\documentclass[aps,prb,preprint,groupedaddress,showpacs]{revtex4-1}

\usepackage{graphicx}

\newcommand{\bnc}{$g$-B$_{3}$N$_{3}$C}
\newcommand{\bnca}{$\alpha$-$g$-B$_{3}$N$_{3}$C}
\newcommand{\bncb}{$\beta$-$g$-B$_{3}$N$_{3}$C}
\newcommand{\ete}{{\it et al.}}

\begin{document}


\title{$g$-B$_{3}$N$_{3}$C: a novel two-dimensional graphite-like material}


\author{J. Y. Li}
\author{D. Q. Gao}
\author{X. N. Niu}
\author{M. S. Si}
\email{sims@lzu.edu.cn}
\author{D. S. Xue}
\email{xueds@lzu.edu.cn}

\affiliation{Key Laboratory for Magnetism and Magnetic Materials of
  the Ministry of Education, Lanzhou University, Lanzhou 730000,
  China}


\date{\today}

\begin{abstract}
 A novel crystalline structure of hybrid monolayer hexagonal boron nitride (BN) and graphene is predicted by
 means of the first-principles calculations. This material can be derived via boron or nitrogen
 atoms substituted by carbon atoms evenly in the graphitic BN with vacancies.
 The corresponding structure is constructed from a BN hexagonal ring linking an additional carbon atom.
 The unit cell is  composed of 7 atoms, 3 of which are boron atoms, 3 are nitrogen atoms, and one is carbon atom.
 It behaves a similar space structure as graphene, which is thus coined as {\bnc}. Two stable
 topological types associated with the carbon bonds formation, i.e., C-N or C-B bonds, are identified.
 Interestingly, distinct ground states of each type, depending on C-N or C-B bonds, and electronic
 band gap as well as magnetic properties within this material have been studied systematically.
 Our work demonstrates practical and efficient access to
 electronic properties of two-dimensional nanostructures providing an approach to tackling open fundamental
 questions in bandgap-engineered devices and spintronics.

\end{abstract}

\pacs{73.22.Pr, 75.70.Ak, 78.67.Wj}

\maketitle


\section{Introduction}

Two-dimensional (2D) nanomaterials, such as graphene and monolayer hexagonal boron nitride (h-BN),
are expected to play a key role in future nanotechnology as well as to provide potential
applications in next-generation electronics. Recently, novel hybrid structures
consisting of a patchwork of BN and C nanodomains (BNC) was synthesized through the
use of a thermal catalytic chemical vapour deposition method \cite{lijie}. This finding
immediately has attracted a great deal of research interest \cite{angel, jeil, marco},
given it demonstrating a hitherto efficient route to tune the band gaps of these 2D materials.

It is well known that the perfect hexagonal and planar structure of BNC largely depends on the well matching
between BN and C domains. However, it is indeed an outstanding challenge as BN and C phases are naturally
immiscible in 2D \cite{lijie}. This explains why Ci {\ete} \cite{lijie} could observe some wrinkles
in atomic force microscopy (AFM) image. Such mutual contradiction mainly originates from the domain boundary
effect and the staggered potentials of B and N atoms in BNC, which doubtlessly affects their
continuous tunable electronic energy gaps. It has been confirmed in the related theoretical calculations
\cite{apl1, apl2, apl3, nano1, jpcc1} where the band gaps behave a strong oscillation feature.

The X-ray photoelectron spectroscopy (XPS) of BNC measured in the work from Ci {\ete} \cite{lijie}
shows two additional types of C bonding configurations, which correspond to the C-B and C-N bonds with the bonding
energies of around 188.4 and 398.1 eV, respectively. This feature means that two inequivalent C bonding types,
i.e., C-B or C-N bonds, must be present in the boundaries of the hybridized BN and C domains, which have a
significant effect on determining the local structures and subsequently vary the electronic properties in
this system.  For example, the transport channels behave a robust characteristic gap when the
topological index changes sign of the valley Hall effect \cite{jeil}. In addition, both Raman D band
at 1,360 cm$^{-1}$ and D$'$  band at 1,620 cm$^{-1}$ are also observed in BNC \cite{lijie}, which were
attributed to the lattice disorder or the finite crystal size. This lattice disorder effect might directly
introduce vacancies to these 2D hexagonal system \cite{lehtinen, sims1, sims2}, which is also true in BNC
as shown in Fig. 2a of Ref. \cite{lijie}.

For the planar BNC structure, although the larger domains would be preferred to decrease the total
domain interfacial energy, the randomly distributed hybrid domains, the immiscible phases as well as the
induced vacancies must result in various complex structures of BNC \cite{pruneda, mazzoni, enyashin, dutta}.
As an immediate consequence is the largely inaccessible synthesis of expected BNC in experiments \cite{aplhan},
which next hinders realization of bandgap-engineered applications in actual devices.
In this endeavour, exploring the structures and electronic properties associated with C bonds formation
in BNC that contain C-N or C-B bonds is an interesting topic that must be addressed before
widespread synthetic applications. Thus, a simple model of BNC where the C bonds play a crucial role
must be considered again. More importantly, a deep theoretical understanding, which originally was concealed
behind the complex hybridized structures, is imperative.

\begin{figure}
\centering
\includegraphics[width=10cm]{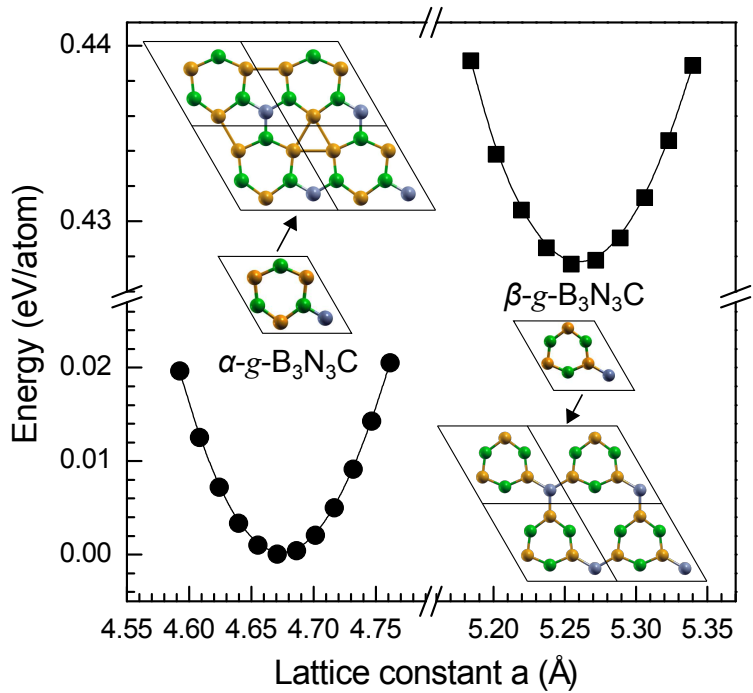}
\caption{The total energy per atom as a function of lattice constant $a$
         for {\bnc}. Yellow, green, and gray balls represent B, N, and C atoms,
          respectively.  Their respective 2$\times$2 supercells are also given nearby.
         }
\label{Fig1}
\end{figure}

Here we report such a simple model of BNC may be just the graphitic B$_{3}$N$_{3}$C ({\bnc}),
which is a perfect 2D monolayer graphite-like structure as shown in Fig. 1. As mentioned above,
C atom can bond to B and N atoms.  Therefore, two topological types of {\bnc} are easily deduced.
One is the {\bnca} related to the the higher bonding energy of C-N bond, while the other one is
the {\bncb} constructed based on the lower bonding energy of C-B bond. It can be seen that such
a material is essentially C-doped graphitic BN (g-BN) with
vacancies \cite{sims1}. The substitution of the N/B atom with a C atom in g-BN with the B/N vacancy will
yield the {$\alpha/\beta$-g-B$_{3}$N$_{3}$C} structure. The interactions among the C atoms and/or vacancies
as well as the C bonding types (C-B and C-N bonds) in {\bnc} significantly alter its electronic properties. To explore
this effect, standard density functional theory with different functionals (see the following discussion)
calculations have been carried out for this predicted material. Remarkably, such material displays two distinct
electronic structure properties: {\bnca} is a semiconductor, while {\bncb} behaves metallic and leads to a magnetic
ground state.

This paper is arranged as follows: In the second section, we present the computational method used in this work,
followed by electronic band gap in {\bnca} and magnetism in {\bncb} in the third section.
We then conclude this paper in the fourth section.

\section{Computational method}

For structural optimization, we employed density functional theory with the generalized gradient
approximation (GGA) of Perdew-Burke-Ernzerhof (PBE) \cite{perdew1}
for the exchange-correlation (XC) potential within the projector
augmented wave method as implemented in {\footnotesize VASP} \cite{kresse1,*kresse2}. An all-electron description, the projector
augmented wave method, is used to describe the electron-ion interaction. The cutoff energy
for plane waves is set to be 500 eV, and the vacuum space is at least 15 {\AA}, which is large
enough to avoid the interaction between periodical images. A 7$\times$7$\times$1 Monkhorst-Pack
grid is used for the sampling of the Brillouin zone during geometry optimization. All the atoms
in the unit cell were allowed to relax, and the convergence of force is set to 0.01 eV/{\AA}.
Additionally, spin polarization is turned on during the relaxation processes.
All other calculations of accurate electronic properties were performed using the full-potential
linearized augmented plane-wave (FLAPW) \cite{madsen} method as implemented in the {\footnotesize WIEN2k} code \cite{blaha}.
It is well known that different XC potentials can lead, depending on the studied materials and properties, to results
which are in very bad agreement with experiment, e.g., for the band gap of semiconductors and insulators which
is severely underestimated or even absent \cite{heyd}.
For this reason, the modified Becke and Johnson (MBJ) \cite{jcp1, prl2} potential in the framework of local-density
approximation (LDA) \cite{kohn, perdew2} is taken to calculate the band gap of {\bnca}, while the magnetism in
{\bncb} is described by the PBE (the standard GGA for materials) potential.

\section{Results and Discussion}

\subsection{Electronic band gap in $\alpha$-{\it g}-B$_{3}$N$_{3}$C}

Figure 1 plots the total energy per atom against the lattice constant $a$ (the lattice constant $c$ is fixed)
for {\bnc}. It can be seen that the total energy of {\bnc} as a function of lattice constant $a$ has a single
minimum, meaning that the geometrical structure would be stable. Particularly, the charge population analysis
reveals that the electron density around the C-N bond in {\bnca} is much higher than that on the C-B bond in
{\bncb}, showing that the C-N bond is relatively strong, which is also consistent with the experimental
results \cite{lijie}. This strong interaction between C and N atoms in {\bnca} directly results in a short
C-N bond length (see following), and can balance the strain of the monolayer graphite-like structure.
Thereby {\bnca} could be a more thermodynamically stable topological phase against {\bncb}, which has
a lower total energy of around 0.43 eV/atom compared with {\bncb}.

\begin{figure}
\centering
\includegraphics[width=8.2cm]{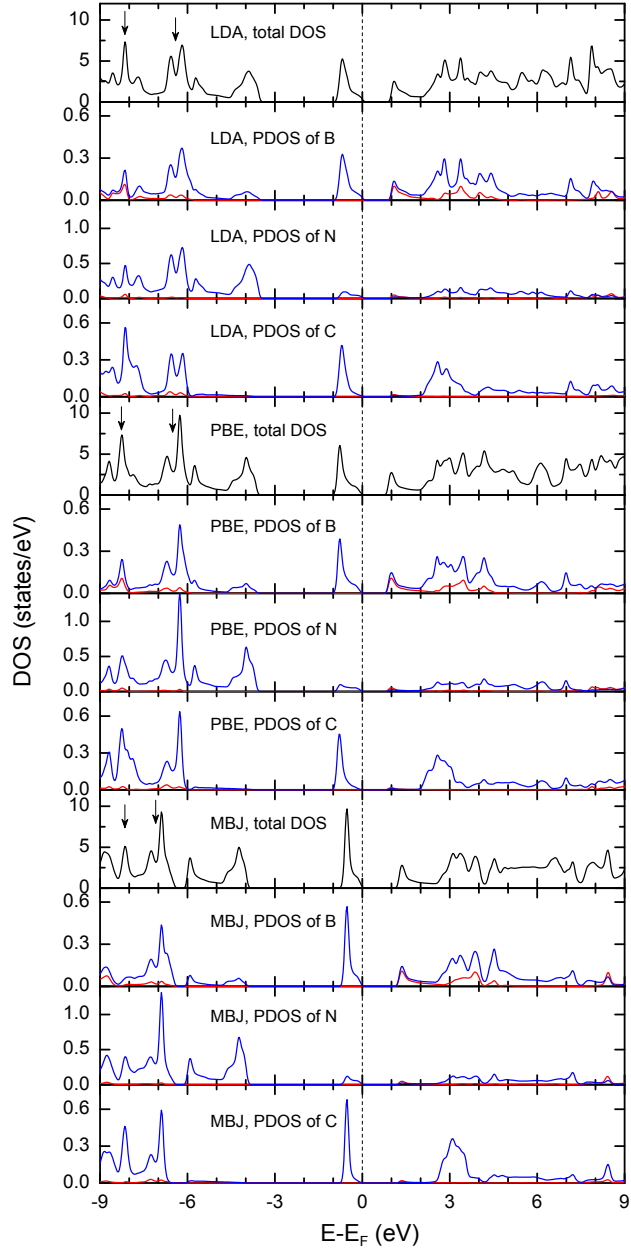}
\caption{Total and partial DOS of {\bnca} for three XC potentials LDA, PBE, and MBJ.
         The vertical dotted line denotes the Fermi level,
         and also indicates end of the fundamental band gap which starts at $E-E_{\rm F}=0$
         eV. The black, blue, and red lines correspond to the total, {\it p}, and {\it s} DOS,
         respectively.
  }
\label{Fig2}
\end{figure}

To explore further the mechanical stability of {\bnca}, the optimized lattice constant $a=4.67$ {\AA} is first
obtained, as depicted in the left panel of Fig. 1. Importantly, the C-N bond length converged to 1.31
{\AA}, considerably reduced from the typical bond lengths 1.37-1.48 {\AA} in the related materials \cite{arxiv}.
This strongly suggests that the nature of the higher binding energy of C-N bond \cite{lijie}, which also
influences the B-N bonding and extends its length. The obtained  B-N bond length is 1.48 {\AA}, which is
slightly bigger than the value of 1.45 {\AA} in h-BN. The calculated partial density of states (PDOS) is shown in
Fig. 2. The valence band is dominated by B {\it p} and C {\it p} states, while the conduction band is only dominated
by B {\it p} states. There one can find that the majority of C {\it p} states to be semicore lying 6-9 eV below the
Fermi level. These states interact with those comprising the valence band with the same symmetry.
As a result there is a small admixture of C {\it p} and N {\it p} states close to the Fermi level.
However, in {\bnca} one finds significant admixture of C {\it p} and N {\it p} states in the semicore
energy window with 6-9 eV below the Fermi level. This suggests that the C {\it p}-N {\it p} interaction in
semicore region contribution to the C-N bonding is significantly more important in {\bnca} than
the C {\it p}-N {\it p} interaction close to the Fermi level.

\begin{figure}
\centering
\includegraphics[width=12cm]{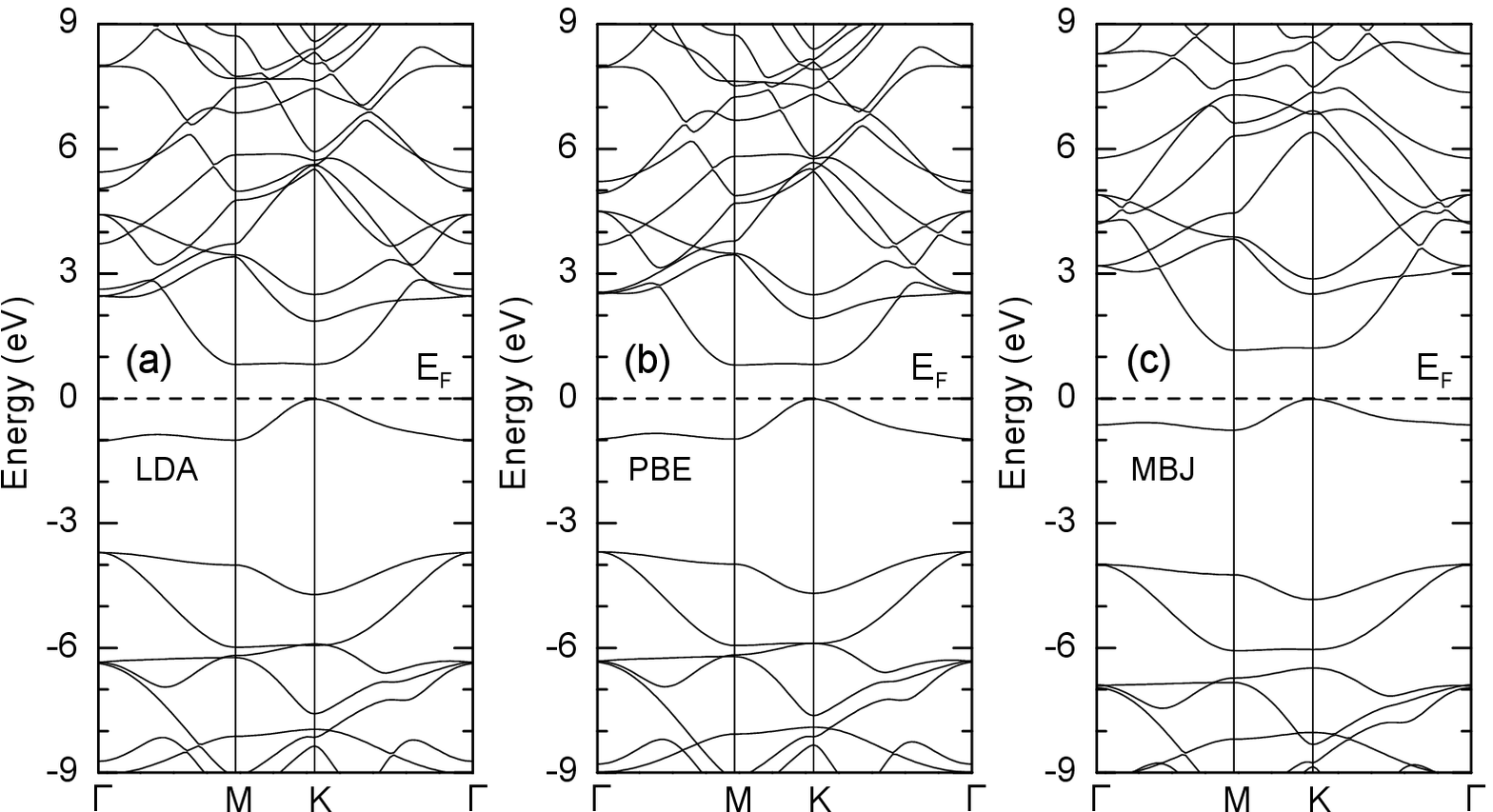}
\caption{Calculated band structures for {\bnca} model with
          the three XC potentials LDA (a), PBE (b), and MBJ (c).
  }
\label{Fig3}
\end{figure}

Now, let us look at the band structures of {\bnca}, as given in Fig. 3. It explicitly demonstrates that all three XC potentials give similar band structures. The highest occupied crystalline
orbitals (HOCOs) are located at the K point of the reciprocal space, while the lowest unoccupied crystalline
orbitals (LUCOs) appear at the M point. This leads to an indirect-band-gap semiconductor. To obtain the band gap
more accurately [see Fig. 3(c)], the MBJ XC functional is used
\cite{jcp1, prl2}. The band gap of {\bnca} within MBJ is obtained to be 1.22 eV, which nicely locates the middle
region between 0.59 and 1.80 eV for the BNC samples with 12.5\% and 50\% C contents, respectively \cite{marco}.
Note that the C content in {\bnca} is 25\%.  In the experiments \cite{lijie}, the absorption edges are redshifted
as the C concentrations increase, which shows a tunable mechanism of optical band gap in actual applications.
By comparing with Ref. \cite{lijie} where BNC with around 65\% C concentration shows an optical bandgap of 1.62
eV, we infer that such a higher energy absorption edge (take into account the band gap of 1.22 eV in our case
with 25\% C concentration) arises from the formation of individual BN and graphene domains. In this way, the
even distribution of C in BNC systems might  serve as a good guide to find alternative solutions to existing
bandgap-engineered applications.  Future researches can test this prediction directly.

\begin{table}
\centering
\caption{\label{tab1}The band gaps (in eV) and the relative band gap correction
         $\cal R$ (\%) of 16 {\it sp} semiconductors and the predicted {\bnca}. The
         theoretical and experimental band gaps of the 16 {\it sp} semiconductors
         are directly taken in \cite{prl2}.}
\begin{ruledtabular}
\begin{tabular}{ccccc}
Solid & LDA  & MBJ  & $\cal R$  & Expt. \\
\hline
C    &  4.11 &  4.93 &  16.6 &  5.48 \\
Si   &  0.47 &  1.17 &  59.8 &  1.17 \\
Ge   &  0.00 &  0.85 & 100.0 &  0.74 \\
LiF  &  8.94 & 12.94 &  30.9 & 14.20 \\
LiCl &  6.06 &  8.64 &  29.9 &  9.40 \\
MgO  &  4.70 &  7.17 &  34.4 &  7.83 \\
ScN  & -0.14 &  0.90 & 115.6 &  0.90 \\
SiC  &  1.35 &  2.28 &  40.8 &  2.40 \\
BN   &  4.39 &  5.85 &  25.0 &  6.25 \\
GaN  &  1.63 &  2.81 &  42.0 &  3.20 \\
GaAs &  0.30 &  1.64 &  81.7 &  1.52 \\
AlP  &  1.46 &  2.32 &  37.1 &  2.45 \\
ZnS  &  1.84 &  3.66 &  49.7 &  3.91 \\
CdS  &  0.86 &  2.66 &  67.7 &  2.42 \\
AlN  &  4.17 &  5.55 &  24.9 &  6.28 \\
ZnO  &  0.75 &  2.68 &  72.0 &  3.44 \\
{\bnca}  &  0.83 &  1.22 &  32.0 &  $-$\\
\end{tabular}
\end{ruledtabular}
\end{table}

In addition, the band gaps based on the LDA and PBE methods are equal
to 0.83 eV [see Figs. 3(a) and (b)]. The relative band gap correction from MBJ with respect to LDA,
 ${\cal R}=(\Delta_{\rm MBJ}-\Delta_{\rm LDA})/\Delta_{\rm MBJ}$
with $\Delta_{\rm MBJ/LDA}$ being the band gap, is about 32{\%}. We can see that our
calculated $\cal R$ for {\bnca} ( shows an excellent agreement with the
16 {\it sp} semiconductors) lies within the range 16.0\%-100.0\% (see Table 1 for details).
The correction value is particularly
very close to that in BN (25\%), GaN (42\%), AlP (37\%), and AlN (25\%). This finding is
not surprised because the listed 4 solids have at least one element close to that of
{\bnca} in the periodic table of elements. It means the similar chemical circumstances in
these materials can be well described by the same XC functionals.
This underlyingly confirms our prediction validity and the {\bnca} might be carried out
experimentally.

From Fig. 2, which shows the DOS of {\bnca}, we can see that the effect of the MBJ potential is
to shift up (with respect to LDA/PBE) the unoccupied B 2{\it s} and 2{\it p} states. Here, three
major differences between the LDA/PBE and MBJ methods can be extracted. (a) Another obvious effect of
MBJ potentials is to shift down the middle of the valence band at around -4.0 eV. (b) The hybridization
of {\it s} and {\it p} states of dominant B 2{\it s} and other atoms 2{\it p} states at the bottom of
the valence band is very strong in MBJ calculations (denoted as down arrows in Fig. 2). (c) The correction
character of {\bnca} is much more
pronounced with the MBJ than with the LDA/PBE, which narrows the valence band just below the Fermi level.
We would like to stress that the MBJ potentials open a band gap of 0.39 eV in the {\bnca} model compared to the result of LDA, which
is consistent with the orbital-dependent potentials principle \cite{prl2}.

\begin{figure}
\centering
\includegraphics[width=10cm]{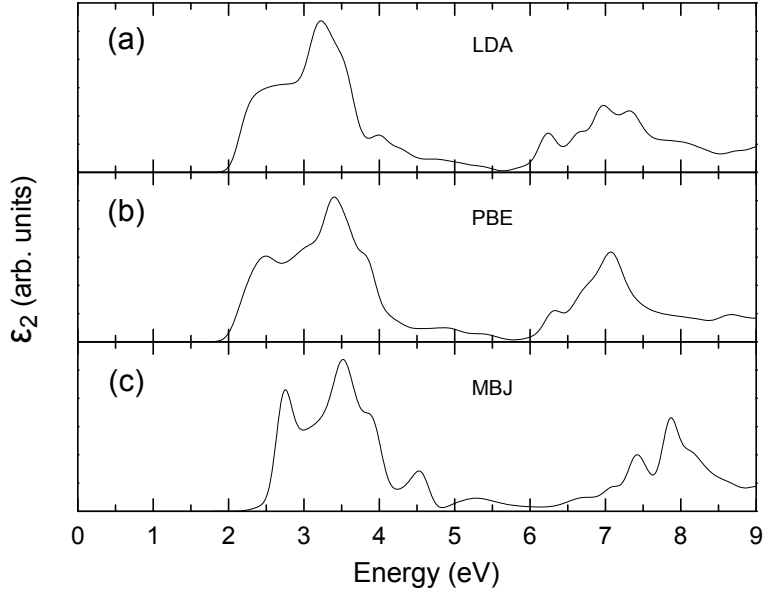}
\caption{The optical absorption expressed by the imaginary part of the
         dielectric tensor $\varepsilon_{2}$ for {\bnca}.
         The imaginary part of the dielectric tensor is averaged for four different directions.
         Three XC potentials including LDA (a), PBE (b), and MBJ (c) are shown in calculations.}
\label{Fig4}
\end{figure}

The optical absorption spectrum for {\bnca} within MBJ potentials [see Fig. 4(c)] shows a big absorption
packet with two adjacent peaks in the range of 2.5-5.0 eV, which originates from the band structure as
shown in Fig. 3(c). As an indirect band gap material, the transition from the K to $\Gamma$ point is very
weak as the momentum conservation rule is not satisfied here, and thus the corresponding
characteristic photoluminescence in the optical absorption spectrum can be negligible. This is also true
in our case {\bnca}. The first peak corresponds to the direct band gap transition at the M point. The
second peak comes from the larger direct gap from higher energy states located at the $\Gamma$ point.
More importantly, these two peaks are not separated distinctly. This feature can be attributed to the
even distribution of C atoms in {\bnca} as mentioned above, which behaves a different formation mechanism
compared with the hybrid BNC in experiment \cite{lijie}.

Correction effects were taken into account by adding a LDA correction potential in MBJ \cite{prl2}.
This important physical effect opens an additional band gap through mimicing very well the behavior
of orbital-dependent potentials and causes a rigid blueshift of the absorption spectrum compared
with the LDA/PBE curves as shown in Fig. 4. This explains the excellent qualitative agreement of the
hybrid exchange potential optical absorption spectrum seen in Fig. 4, due to a compensation of
significant errors within the standard DFT methods.

\subsection{Magnetism in $\beta$-{\it g}-B$_{3}$N$_{3}$C}

The predicted structure of {\bncb} is shown in the right panel of Fig. 1.
The lattice constant of {\bncb} is obtained to be 5.26 {\AA} as shown in Fig. 1.
The results show that the equilibrium value $d_{\rm BC}=1.52$ {\AA}, which is close to the
value of graphite-like BC$_{3}$, namely, 1.55 {\AA} \cite{prb1}. It is to be noticed that all the equilibrium values
$d_{\rm BN}$ in {\bnca} are equal to 1.42 {\AA}, which is slightly less than the value of 1.45 {\AA} in
pristine BN sheet. This implies the stronger B-N bonds formed in {\bncb}.
Our calculations show that the {\bncb} leads to a ground state
with a magnetic moment of 0.68 $\mu_{\rm B}$. The nonmagnetic state is 0.07 eV higher than this ground state.

\begin{figure}
\centering
\includegraphics[width=11cm]{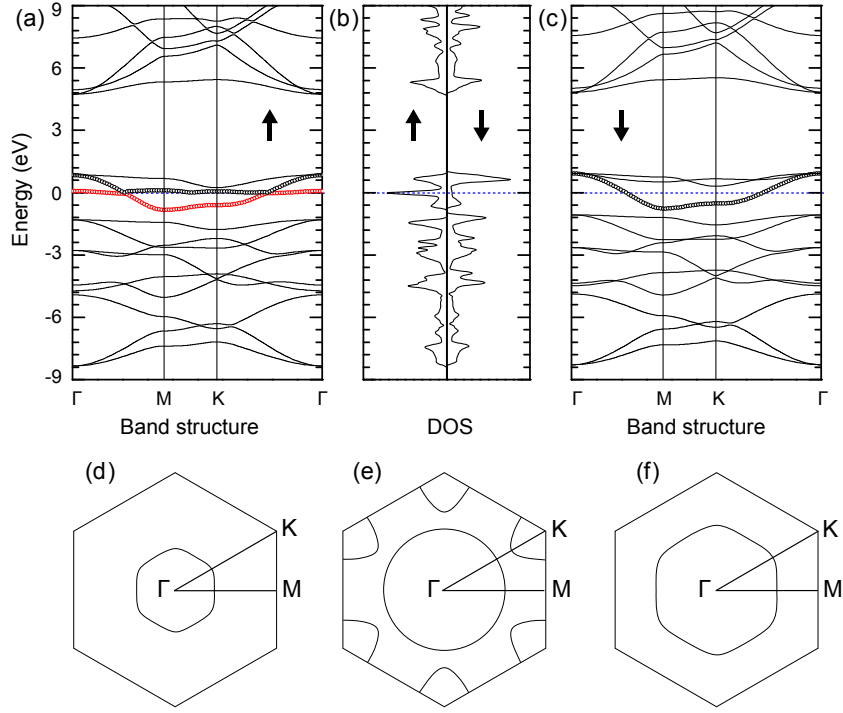}
\caption{(a)-(c) Band structures and spin-resolved TDOS for
         {\bncb}. The dotted line indicates the Fermi level. The arrow denotes
         the spin polarization direction: up for spin up and down for spin down. (d)-(f)
         Fermi surfaces drawn in the first Brillouin zone and the corresponding high-symmetry
         points. (d) and (e) for spin up and (f) for spin down.
  }
\label{Fig5}
\end{figure}

From the band structures we see that although both the pristine BN sheet and graphene are nonmagnetic,
the {\bncb} model can be spin polarized. It is necessary to discuss magnetism in more detail from its
electronic structures. Figures 5(a)-(c) present the band structure and spin-resolved total density of state (TDOS).
Remarkably, two bands cross the Fermi level [black and red circles in Fig. 5(a)], and make the Fermi energy level
occupied completely along the entire high-symmetry lines. In contrast, the spin-down one does not
possess such a strongly localization feature, just one band (black circles) crosses the Fermi level
monolithically as shown in Fig. 5(c). The calculated magnetic moment in {\bncb} should originate
from this asymmetric spin-dependent localization. The corresponding strong spin splitting can be further
confirmed from TDOS as shown in Fig. 5(b). In addition, close examination of
the top valence bands [see Fig. 6(a)] indicates that the strong localization mainly comes from the
2{\it p} atomic orbitals of C and N atoms. Here we also show that the Fermi surfaces of {\bncb}
in the first Brillouin zone. The unique feature of Fermi surface almost parallel to the high-symmetry
lines [see Figs. 5(d)-(f)] is the direct manifestation of the bands near the Fermi level in Figs.
5(a) and 5(c).

It should be notice that such magnetism is induced without transition metals and without external
perturbations, so that {\bncb} behaves as the first, theoretically predicted, metal-free magnetic
material in hybrid BNC system. Apparently, our finding points out a new direction for further related experimental
investigations in spintronics.

\begin{figure}
\centering
\includegraphics[width=12cm]{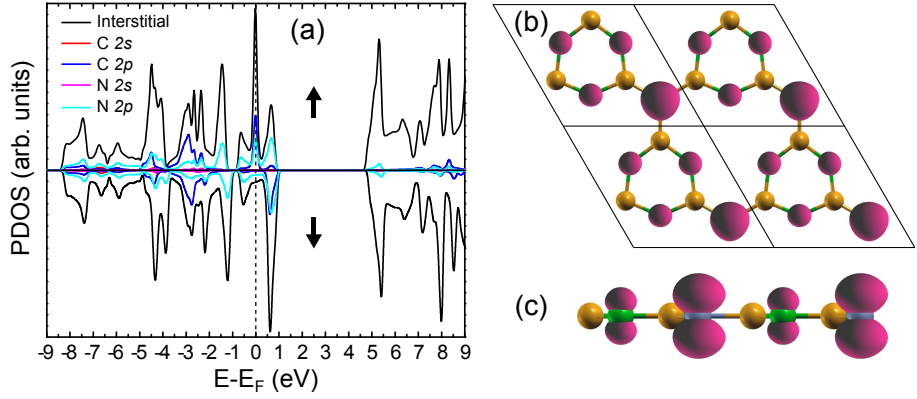}
\caption{(a) PDOS on the interstitial region and on
         the 2{\it s} and 2{\it p} orbitals of C and N atoms in {\bncb}. (b) The
         3D iso-surface plot of spin density for the 2$\times$2 supercell at
         the value of 0.04 e/{\AA}$^{3}$: (c) A side view of spin density corresponding to (b).
  }
\label{Fig6}
\end{figure}

We now address the possibility to magnetism origin in {\bncb}. Based on the orbital-resolved density
of state as shown in Fig. 6(a), the magnetic moment are mainly ascribed to the 2{\it p} orbitals of
C and three N atoms. The spin polarization of C atom offers a magnetic moment of 0.18 $\mu_{\rm B}$.
0.18 $\mu_{\rm B}$ of total magnetic moment are shared equally by the 2{\it p} orbitals of three N atoms.
The remaining magnetic moments distribute evenly in the interstitial region among N atoms. This
is conceivable, because {\bncb}, if compared with other BNC systems, has large interspaces between
N atoms [see the 2$\times$2 supercell in Fig. 1 or Fig. 6(b)]. The remarkable feature is that the equilibrium surface
density of {\bncb} is around 1.27 times larger than {\bnca}. Owing to the large interspaces between atoms, the
hydrogen storage in {\bncb} may be expected \cite{hydrogen}. The detailed description of hydrogen storage related
to {\bncb} is beyond the scope of this work.

Concerning the special carbon atom in {\bncb}, the counting of four valence electrons is as follows: Three
electrons participate in the
$sp^{2}$ hybrid orbital, which forms a planar structure. The remaining one electron is then redistributed in
the whole unit cell due to the enhanced B-N covalent bond (with shorter bond length compared with the value in
pristine h-BN), which makes the magnetic properties more
complicated. From the fourth electron, only 18 percent of the electron still fill the $\pi$-orbital of C atom
and contribute a magnetic moment of 0.18 $\mu_{\rm B}$. This is in fairly good agreement with
theoretical description of the $\pi$-orbital state to the local magnetic moment $\sim$0.3 $\mu_{\rm B}$
in the graphene-like systems \cite{prl1, prb2}. Around 32\% of the fourth electron mainly reside at the
interstitial region, which play a crucial role in {\bncb} as: (a) Enhance the B-N covalent bond.
(b) Provide the main interstitial magnetic moment of 0.32 $\mu_{\rm B}$. (c) Promote the 2$p_{z}$ of N
atom spin polarized slightly with a magnetic moment of 0.06 $\mu_{\rm B}$ per N atom. The remaining of the fourth
electron act as the conduction electrons and make the system metallic, which dominate the mechanism of
ferromagnetic ordering in {\bncb}.  In the case of {\bncb}, we can
see that the electron spin at the localized $\pi$-orbital state of C and N atoms, as well as the
interstitial region compel two energy bands localized strictly along the entire high-symmetry lines, i.e.,
the $\Gamma$-M-K-$\Gamma$ line [see Fig. 5(a)]. Thus, the RKKY interaction \cite{prb3, prb4} among
the magnetic sites through the residual conduction electrons forms a spin ordering in these orbitals,
which is the physical origin of ferromagnetism in {\bncb}. Figures 6(b) and 6(c) respectively plot the
top and side views of the 3D iso-surfaces for net magnetic charge density in {\it xy} plane. This finding is
insightful and three major points deserve comment: (a) The C site is more spin polarized as compared with
each N site. (b) The dumbbell-like magnetic moment distribution along {\it z} direction implies the 2$p_{z}$
orbital becomes partially filled with one spin-up electron. (c) The induced moments are ferromagnetic coupled
between the N and C sites based on the RKKY exchange interaction model as mentioned above.

\section{Conclusions}

In summary, we have predicted a novel crystalline material {\bnc}, which displays two distinct electronic properties where the selective bonding type of C atom is a key parameter for future industrial processes.
{\bnca} is a  semiconductor, while {\bncb} behaves metallic and holds a magnetic moment of 0.68 $\mu_{\rm B}$.
Importantly, compared with the hybrid BNC, {\bnc} is proposed to have a simple structure, which can be
applied in various fields due to its unique properties.

\begin{acknowledgments}
This work was supported by the National Basic Research Program of
China under No. 2012CB933101. This work was also supported by the
National Science Foundation of China (NSFC) under No. 10804038,
11034004, and 50925103, and the Fundamental Research Fund for the
Central Universities and Physics and Mathematics of Lanzhou University.
We acknowledge part of the work as done on National Supercomputing Center in Shenzhen.
\end{acknowledgments}

%

\end{document}